\documentclass[%
 reprint,
 amsmath,amssymb,
 aps,
]{revtex4-2}

\usepackage{tikz}
\usetikzlibrary{shapes.geometric, arrows}

\usepackage{listings}
\usepackage{xcolor}
\usepackage{booktabs} 
\usepackage{graphicx} 
\usepackage{hyperref} 
\usepackage{makecell} 

\definecolor{codebg}{rgb}{0.95,0.95,0.95} 
\definecolor{keywords}{rgb}{0,0,1} 
\definecolor{strings}{rgb}{0.56,0,0} 
\definecolor{comments}{rgb}{0.3,0.5,0.3} 
\definecolor{numbers}{rgb}{0.5,0.5,0.5} 
\definecolor{identifiers}{rgb}{0.1,0.1,0.1} 
\definecolor{operators}{rgb}{0.4,0.4,0.4} 

\tikzstyle{block} = [rectangle, draw, fill=gray!20, text width=7em, text centered, rounded corners, minimum height=2em, font=\small]
\tikzstyle{line} = [draw, -latex']
\tikzstyle{dashedline} = [draw, -latex', dashed]  

\usepackage{graphicx}
\usepackage{dcolumn}
\usepackage{bm}

\usepackage{xcolor}

\begin{document}

\preprint{APS/123-QED}

\title{Fine-Tuning Universal Machine-Learned Interatomic Potentials: A Tutorial on Methods and Applications}
\author{Xiaoqing Liu}
\affiliation{School of Mathematical Sciences, Shanghai Jiao Tong University, Shanghai 200240, China}
\affiliation{Shanghai Jiao Tong University-Chongqing Institute of Artificial Intelligence, Chongqing 401329, China}

\author{Kehan Zeng}
\affiliation{Shanghai Jiao Tong University-Chongqing Institute of Artificial Intelligence, Chongqing 401329, China}

\author{Zedong Luo}
\affiliation{Shanghai Jiao Tong University-Chongqing Institute of Artificial Intelligence, Chongqing 401329, China}

\author{Yangshuai Wang}
\email{yswang@nus.edu.sg}
\affiliation{Department of Mathematics, National University of Singapore, 10 Lower Kent Ridge Road, Singapore}

\author{Teng Zhao}
\email{zhaoteng\_sjtu@sjtu.edu.cn}
\affiliation{Institute of Natural Sciences, MOE-LSC, and Shanghai National Center for Applied Mathematics, Shanghai Jiao Tong University, Shanghai 200240, China}
\affiliation{Shanghai Jiao Tong University-Chongqing Institute of Artificial Intelligence, Chongqing 401329, China}

\author{Zhenli Xu}
\email{xuzl@sjtu.edu.cn}
\affiliation{School of Mathematical Sciences, MOE-LSC, CMA-Shanghai and Shanghai Center for Applied Mathematics, Shanghai Jiao Tong University, Shanghai, 200240, China}


\date{\today}

\begin{abstract}
    Universal machine-learned interatomic potentials (U-MLIPs) have demonstrated broad applicability across diverse atomistic systems but often require fine-tuning to achieve task-specific accuracy. While the number of available U-MLIPs and their fine-tuning applications is rapidly expanding, there remains a lack of systematic guidance on how to effectively fine-tune these models. This tutorial provides a comprehensive, step-by-step guide to fine-tuning U-MLIPs for computational materials modeling. Using the recently released MACE-MP-0 as a representative foundation model, we illustrate the full workflow of dataset preparation, hyperparameter selection, model training, and validation. Beyond methodological guidance, we conduct systematic case studies on solid-state electrolytes, stacking fault defects in metals, semiconductors, solid–liquid interfacial interactions in low-dimensional systems, and more complicated heterointerfaces. These examples demonstrate that fine-tuning substantially improves predictive accuracy while maintaining affordable computational cost, accelerates training convergence, enhances out-of-distribution generalization, and achieves superior data efficiency. Remarkably, fine-tuned foundation models can even capture aspects of long-range physics without explicit corrections. Together, these results highlight that fine-tuning not only provides a practical recipe for applying U-MLIPs, but also offers new insights into their physical fidelity and potential for advancing large-scale atomistic simulations. To support practical applications, we include code examples that enable researchers, particularly those new to the field, to efficiently incorporate fine-tuned U-MLIPs into their workflows.
\end{abstract}

\maketitle


\section{INTRODUCTION}
\label{sec:intro}

Machine-learned interatomic potentials (MLIPs) have become an essential tool in materials modeling, offering near-density functional theory (DFT) accuracy at significantly reduced computational cost~\cite{behler2007generalized, bartok2010gaussian, shapeev2016moment, wang2018deepmd, DrautzACE, batatia2022mace, schutt2017schnet, batzner20223, cheng2024cartesian, musaelian2023learning, smith2017ani, bochkarev2024graph, thompson2015spectral, xie2023ultra}. A broad range of MLIP frameworks have been developed in recent years, including neural network potentials~\cite{behler2007generalized, wang2018deepmd}, kernel-based approaches~\cite{bartok2010gaussian, DrautzACE}, and equivariant graph neural networks~\cite{schutt2017schnet, batatia2022mace}, each demonstrating excellent performance across diverse materials systems. These models enable accurate and large-scale atomistic simulations, and their integration with simulation platforms such as ASE~\cite{larsen2017atomic} and LAMMPS~\cite{thompson2022lammps} has facilitated their widespread adoption in the field. For comprehensive reviews on MLIPs, the reader is referred to Refs.~\cite{unke2021machine, botu2017machine, jacobs2025practical, musil2021physics, poltavsky2021machine}.

Traditionally, most MLIPs are trained on narrowly defined chemical or structural domains tailored to specific applications. While such models can achieve high accuracy within their training scope, they often lack transferability to systems beyond that domain. In response, there has been growing interest in a new class of models, often referred to as general-purpose, foundation, or universal MLIPs (U-MLIPs)~\cite{batatia2023foundation, deng2023chgnet, merchant2023scaling, zhang2024dpa, choudhary2023unified, chen2022universal}. Inspired by foundation models in natural language processing and computer vision~\cite{awais2025foundation}, these models are pre-trained on large and diverse materials datasets~\cite{chanussot2021open, bowman2022md17, barroso2024open}, aiming to provide broad coverage across the periodic table and structural configuration space. They offer strong baseline performance and could be readily applied to new systems with minimal additional training (i.e., fine-tuning), substantially lowering the barrier to entry for MLIP applications in materials science. In Table~\ref{tab:mlip_models}, we summarize representative U-MLIPs, along with their publication year, underlying architectures, and training data characteristics.

\begin{table*}[t]  
    \centering
    \caption{List of the representative U-MLIPs.}
    \label{tab:mlip_models}
    \renewcommand{\arraystretch}{1.3} 
    \setlength{\tabcolsep}{4pt} 
    \begin{tabular}{c c c c}
        \toprule
        \textbf{Model Name} & \textbf{Year} & \textbf{Architecture} & \textbf{Data Description} \\
        \midrule
        M3GNET \cite{chen2022universal} & 2022 & SchNet~\cite{schutt2018schnet} & Materials Project (MP), 89 elements, 62783 compounds  \\
        CHGNet \cite{deng2023chgnet} & 2023 & GNN + Charge & MP + Trajectory, 89 elements, 146000 compounds \\
        ALIGNN-FF \cite{choudhary2023unified} & 2023 & Line GNN & JARVIS-DFT, 72708
compounds \\
        PFP (Matlantis) \cite{takamoto2023towards} & 2023 & Tensorial GNN  & Custom, $\approx$10 million
configurations \\ 
        GNoME \cite{merchant2023scaling} & 2023 & NequIP~\cite{batzner20223} & MP +
Custom, $\approx$89 million
configurations \\
        DPA-1 \cite{zhang2024dpa}, DPA-2 \cite{zhang2024dpa} & 2023, 2024 & DeepMD~\cite{wang2018deepmd} & Open-access datasets including Alloy, OC2M, and others.  \\
        {\bf MACE-MP-0} \cite{batatia2023foundation} & 2024 & MACE~\cite{batatia2022mace} & Same data as used to build CHGNet \\
        SevenNet-0 \cite{park2024scalable} & 2024 & NequIP~\cite{batzner20223} & Same data as used to build CHGNet \\
        MatterSim \cite{yang2024mattersim} & 2024 & Graph transformer &  MP+Alexandria+Custom, $\approx$17 million
configurations \\
        EquiformerV2-OMAT24 \cite{barroso2024open} & 2024 & EquiformerV2 \cite{liao2023equiformerv2} & MP+Alexandria+Custom, $\approx$118 million
configurations \\
        \bottomrule
    \end{tabular}
\end{table*}

Among existing U-MLIPs, MACE-MP-0 stands out as a near state-of-the-art model, demonstrating good accuracy and robust stability in molecular dynamics simulations across a wide range of applications~\cite{batatia2023foundation}. It serves as the central focus of this work. MACE-MP-0 is built upon the MACE architecture~\cite{batatia2022mace}, which integrates the atomic cluster expansion (ACE) formalism~\cite{DrautzACE, ACECompleteness} with efficient tensor decomposition techniques~\cite{darby2023tensor} and higher-order equivariant message passing. Leveraging this design, MACE-MP-0 is pre-trained on a large dataset from the Materials Project, and achieves strong performance across diverse materials benchmarks. Several subsequent variants of the model have been introduced recently to address specific issues such as repulsion, phonons, and high-pressure stability, further broadening its applicability.

While MACE-MP-0 and its varaints have demonstrated robust performance in molecular dynamics simulations~\cite{batatia2023foundation}, its predictive accuracy—like that of other U-MLIPs—remains limited for certain tasks, particularly in modeling mechanical properties such as elastic constants and stacking fault energies in elemental alloys~\cite{li2024extendable}. This underscores the need for fine-tuning, which adapts pre-trained models to specific systems by refining their parameters using high-fidelity computational or experimental data. Although the fine-tuning of atomistic foundation models has garnered growing attention, comprehensive and systematic studies remain limited, especially practical tutorials that guide beginners through the fine-tuning process. Recent works have started addressing this gap~\cite{focassio2024performance, deng2024overcoming, yu2024systematic, pyzer2025foundation, shuang2025universal, du2025universal, lee2025accelerating, niblett2024transferability, casillas2024evaluating}, and continued efforts are expected to clarify best practices across various application scenarios. While the number of available U-MLIPs and their fine-tuning applications is rapidly expanding, there remains a lack of systematic guidance on how to effectively fine-tune these models, which serves as the primary motivation of this tutorial.

Beyond the pursuit of higher accuracy, a central motivation for fine-tuning in the context of MLIPs lies in a fundamental question: given a fixed dataset, should one fine-tune an existing foundation model or train a new model from scratch? In practice, fine-tuning is often more efficient in terms of training time and data usage, and in many cases yields comparable or even superior accuracy. While foundation models, pre-trained on large and diverse datasets, effectively capture general chemical and structural patterns, they often lack the precision required for specific systems. Fine-tuning addresses this limitation by incorporating system-specific data, thereby enhancing accuracy with significantly lower data and computational costs. Moreover, fine-tuned models typically converge faster due to the advantage of informed initialization. Therefore, when a suitable foundation model is available, fine-tuning presents a compelling alternative to training-from-scratch. 

This tutorial provides a practical guide to fine-tuning U-MLIPs, aiming to support researchers—especially those new to the field—by filling the current gap in systematic guidance for fine-tuning foundation models. The goal is to provide step-by-step instructions for preparing data, training models, selecting hyperparameters, and applying the resulting potentials in various atomistic simulations. To keep the presentation accessible, theoretical details are kept to a minimum, with references provided for readers seeking a deeper mathematical and physical understanding. The methods introduced here are demonstrated using widely used tools such as ASE, LAMMPS, and the Random Batch Molecular Dynamics (RBMD) interface~\cite{gao2024rbmd}. MACE-MP-0 is selected as the main example due to its broad applicability, high accuracy across benchmark tasks, and demonstrated stability in molecular dynamics simulations~\cite{batatia2023foundation}.

Beyond methodological guidance, we conduct extensive numerical experiments that yield several key insights. Fine-tuning consistently improves predictive accuracy and accelerates convergence while maintaining affordable computational cost. It enhances out-of-distribution generalization and achieves higher data efficiency compared to training from scratch, even when using reduced datasets. Interestingly, fine-tuned foundation models sometimes show improved accuracy even for cases involving long-range effects, suggesting that certain aspects of such interactions may already be partially captured during pre-training. Taken together, these findings show that fine-tuning not only offers a practical recipe for applying U-MLIPs, but also provides deeper insights into their physical fidelity and their potential to advance large-scale atomistic simulations.

It is worth noting that the fine-tuning of MACE-MP-0 can be efficiently carried out using the RBMD package, which enables large-scale particle simulations at the nano- and microscale. In contrast to conventional molecular dynamics frameworks, RBMD leverages random batch algorithms~\cite{RBM_Jin, RBE, RBE_2, IRBE} to efficiently handle nonbonded interactions, supporting simulations of up to 10 million particles on a single GPU with a CPU core~\cite{gao2024rbmd}. Moving forward, we aim to integrate the RBMD platform with MACE-based MLIPs for more practical and large-scale simulations. To support this integration, an open-access platform is available at \url{https://www.randbatch.com/rbmd}, where further examples and implementation scripts for realistic simulation scenarios will be released in the future.

The remainder of this tutorial is organized as follows. In Section~\ref{sec:methods}, we introduce the overall fine-tuning workflow for U-MLIPs, including a brief overview of the MACE-MP-0 framework, data preparation, potential training, and model validation. Section~\ref{sec:examples} presents six illustrative applications of fine-tuning U-MLIPs: (i) force accuracy evaluation in the solid-state electrolyte Li$_{10}$GeP$_2$S$_{12}$ (LGPS), (ii) stacking fault energetics in body-centered cubic molybdenum (Mo), (iii) generalization tests on out-of-distribution in semicondutor silicon, (iv) surface interactions at the graphene–water interface, (v) molecule–oxide interactions, and (vi) solid–solid interfaces, the latter two designed to assess the ability of fine-tuned models to capture long-range effects. Finally, Section~\ref{sec:summary} provides a summary and outlook on the development and fine-tuning of U-MLIP models.

\section{METHODS}
\label{sec:methods}

\begin{figure}
\centering
\includegraphics[height=12cm]{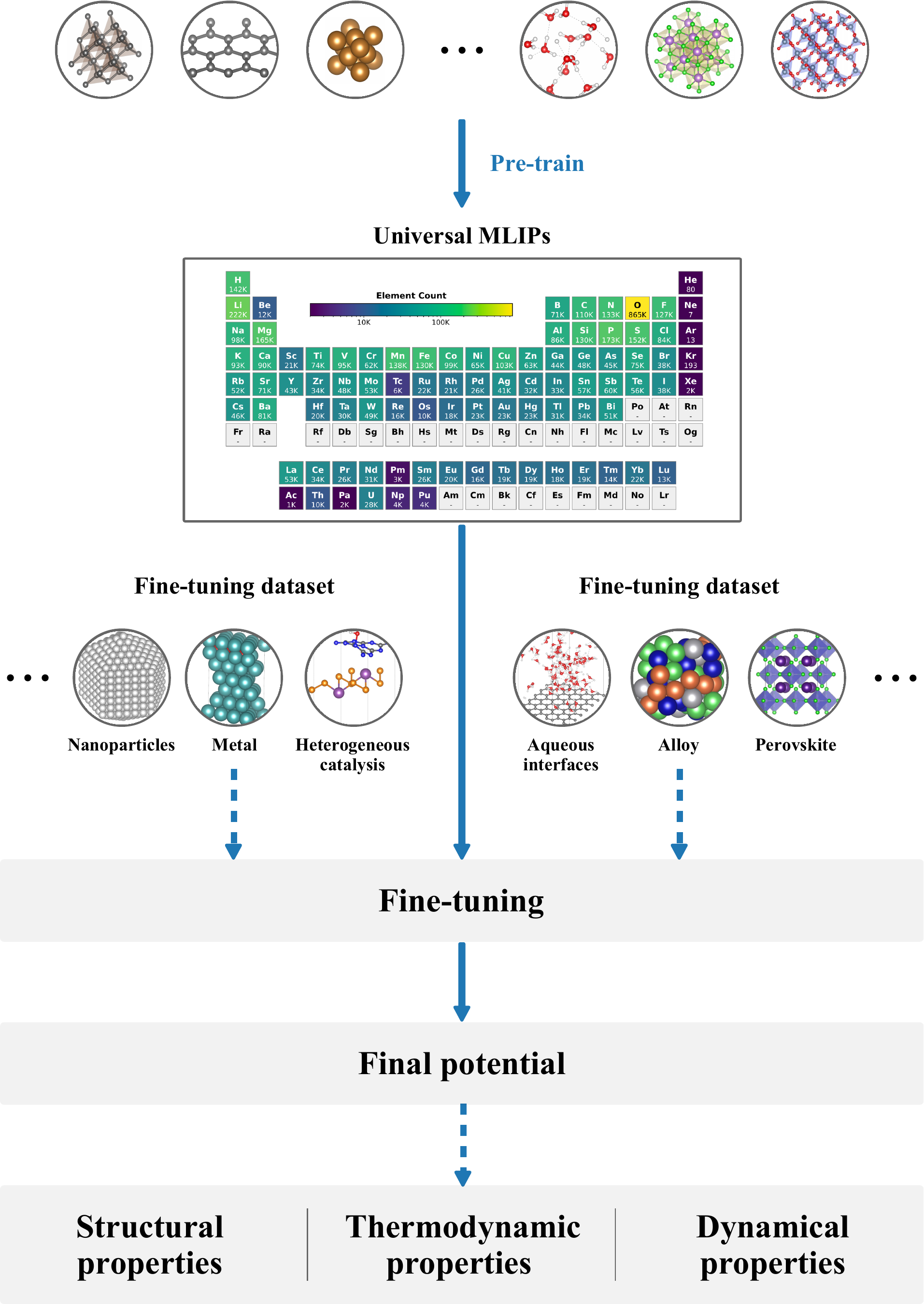}
\caption{Overview of the fine-tuning workflow}
\label{fig:workflow}
\end{figure}

In this section, we introduce the overall procedure for fine-tuning U-MLIPs, using MACE-MP-0 as a representative model. An overview of the fine-tuning workflow is shown in Figure~\ref{fig:workflow}. The section is organized as follows: Section~\ref{sec:sub:mace} briefly describes the MACE-MP-0 model, common fine-tuning practices, and recent implementation updates. Dataset construction, a critical component of MLIP development, is discussed in Section~\ref{sec:sub:data}. Section~\ref{sec:sub:train} outlines the fine-tuning process, including hyperparameter choices. Model validation is addressed in Section~\ref{sec:sub:validation}, which is a prerequisite for reliable molecular dynamics simulations.

The fine-tuning procedure is supported by the official MACE repository:
\url{https://github.com/ACEsuit/mace}.
Installation instructions are also available in the tutorial repository:
\url{https://github.com/John2021-hub/mace-ft-tutorial.git}. The RBMD interface~\cite{gao2024rbmd} supports both fine-tuning and molecular dynamics simulations with MACE-based MLIPs.

\subsection{The MACE-MP-0 Model: Architecture, Fine-Tuning, and Recent Developments}
\label{sec:sub:mace}

We begin by reviewing the MACE-MP-0 model in Section~\ref{sec:sub:pre-trained}. The fine-tuning procedures introduced here are broadly applicable to other U-MLIPs with only minor adjustments.

\subsubsection{Overview of the MACE-MP-0 Model}
\label{sec:sub:pre-trained}

Before introducing U-MLIPs, we briefly recall the concept of MLIPs. These models map atomic positions and chemical elements to the potential energy of a given atomic system. The MACE architecture~\cite{batatia2022mace} is a representative example, extending the Atomic Cluster Expansion (ACE) framework~\cite{witt2023acepotentials, DrautzACE, ACECompleteness}, which was originally developed for MLIPs and has since found applications beyond atomistic modeling~\cite{ACEHam, wang2024theoretical, torabi2024surrogate, wang2025many}. MACE is an equivariant message-passing graph tensor network~\cite{vignac2020building, maron2018invariant}, where many-body atomic geometry is encoded layer by layer. Messages are built from linear combinations of tensor product bases derived from two-body permutation-invariant polynomials in a spherical basis~\cite{DrautzACE, ACECompleteness}. Equivariance is maintained through tensor contractions with irreducible representations and generalized Clebsch–Gordan coefficients~\cite{luo2024enabling, darby2023tensor}. The final output represents the per-atom contribution to the total potential energy. Further architectural details are provided in Ref.~\cite{batatia2022mace}.

The foundation model MACE-MP-0, trained on the Materials Project dataset~\cite{jain2013commentary}, has demonstrated strong performance in a range of applications~\cite{batatia2023foundation}, particularly in enabling stable molecular dynamics simulations across diverse chemical environments. It is released in three variants, differentiated by increasing levels of message-passing equivariance ($L = 0, 1, 2$). A subsequent revision, MACE-MP-0b, introduced minor improvements such as refined pairwise repulsion and improved treatment of isolated atoms. Further variants have been developed to address specific limitations: MACE-MP-0b2 enhances stability under high pressure, MACE-MP-0b3 corrects phonon behavior, and MACE-MPA-0 and MACE-OMAT-0 extend training coverage. All models are publicly available at
\url{https://github.com/ACEsuit/mace-mp/}. In this work, we adopt MACE-MP-0b3 as the base model for fine-tuning. For simplicity, we continue to refer to it as MACE-MP-0 unless otherwise stated.

Despite the general robustness of these models, stable MD simulations do not guarantee accurate predictions of physical properties. Fine-tuning remains essential for adapting U-MLIPs to specific systems. However, to the best of our knowledge, there is currently no systematic tutorial guiding users through the fine-tuning of U-MLIPs for practical applications. Addressing this gap serves as the central motivation for the present work.

\subsubsection{Fine-Tuning Strategies}
\label{sec:sub:ft}

Fine-tuning pre-trained models is widely adopted to improve accuracy and transferability in various domains, including image analysis~\cite{tajbakhsh2016convolutional, jordan2015machine} and natural language processing~\cite{zhang2022fine, devlin2019bert, lee2020biobert}. In molecular modeling, fine-tuning typically follows a predictor–corrector paradigm: a U-MLIP serves as an initial predictor that provides a robust baseline for atomistic simulations. The model is then fine-tuned on a system-specific dataset, allowing it to correct and adapt to the target system with improved accuracy.

Multi-head replay fine-tuning extends this idea by attaching multiple task-specific output heads to a shared pre-trained backbone~\cite{kim2024hydra}. Each head is tailored to a different target system or property, while the backbone captures general representations. By partially freezing the backbone, this approach enables efficient training across multiple tasks and reduces redundancy. This design also mitigates catastrophic forgetting by preserving the performance of previously trained heads while adapting to new tasks. An implementation of this method is available at \url{https://github.com/ACEsuit/mace/tree/main}. 

From an optimization perspective, existing fine-tuning strategies for U-MLIPs can be broadly categorized into:
(i) Full-parameter fine-tuning, where all backbone and head parameters are updated~\cite{kaur2025data}. This maximizes task-specific adaptability but incurs higher computational cost and greater risk of overfitting, especially when the target dataset is small.
(ii) Partial freezing, where early layers or invariant feature extractors are kept fixed while only later layers or heads are updated. This reduces the number of trainable parameters, leading to faster convergence and better knowledge retention from the pre-trained model, but may limit adaptation to systems with strong distribution shifts~\cite{radova2025fine}.
(iii) Lightweight adaptation layers (e.g., low-rank adapters or bias tuning), which introduce a small set of trainable parameters on top of the frozen backbone~\cite{wangelora}. This approach offers a favorable trade-off between efficiency and accuracy, and has shown promise in retaining generalization while improving data efficiency.

In molecular modeling, the choice of strategy should be guided by the similarity between the pre-training and target domains, the size and diversity of the fine-tuning dataset, and the computational budget. A comprehensive benchmark to determine the optimal strategy under different scenarios is still lacking, but future work will aim to provide such guidance.

\subsubsection{Recent Updates}
\label{sec:sub:sub:updates}

This section briefly introduces two recent updates relevant to U-MLIP deployment: the fast equivariant kernel \textsf{cuEquivariance}, and an implementation of long-range dispersion corrections with efficient neighbor list handling. These tools are integrated into our examples (cf.~Section~\ref{sec:examples}) to demonstrate the fine-tuning and application of the MACE-MP-0 model in practical simulations.

\paragraph{The fast equivariant kernel \textsf{cuEquivariance}.}

Equivariance, referring to the preservation of pre-defined physical symmetry under group operations such as rotations and translations, is a key property of physically meaningful interatomic potentials. Models that are naturally equivariant are typically more data-efficient, posses better physicality and exhibit improved generalization.

\textsf{cuEquivariance} is an NVIDIA-developed library for building high-performance equivariant neural networks based on segmented tensor products within easily accessible Python interface. It provides a flexible API and optimized CUDA kernels, with bindings for both PyTorch and JAX. The code is available at
\url{https://github.com/NVIDIA/cuEquivariance}.

Fine-tuned MACE-MP-0 models (cf. Section~\ref{sec:sub:train}) can be integrated with \textsf{cuEquivariance} in Python-based simulation environments such as ASE~\cite{larsen2017atomic}, enabling significantly faster inference—typically by a factor of 3 to 10. Example usage and performance benchmarks are provided in Section~\ref{sec:sub:dp_data}. The following code snippet illustrates how to activate the fast kernel via a simple keyword:

\begin{lstlisting}[caption=Enable cuEquivariance kernel,label=lst:cu_eq]
from mace.calculators import MACECalculator
mace_calc = MACECalculator(model_paths="MACE.model", enable_cueq=True, device="cuda")
\end{lstlisting}

As reported by developers, the jax implementation is expected to have at least extra few factors of speed up further. 

\paragraph{Long-range dispersion corrections with efficient neighbor list handling.}

Frequent neighbor list updates can introduce significant computational overhead, particularly when large cutoffs are used for dispersion corrections such as DFT-D3~\cite{moellmann2014dft}. To mitigate this, a simple yet effective neighbor list implementation with a skin layer is provided in ASE. This approach reduces the frequency of neighbor list reconstruction and improves overall computational efficiency.

The implementation allow similar update strategy in LAMMPS and enables the reuse of neighbor lists across multiple simulation steps. An example configuration is shown below:

\begin{lstlisting}[caption=Enable neighbor list reuse,label=lst:dft_d3]
calc = TorchDFTD3Calculator(atoms=atoms, device="cuda", damping="bj",
    every=2, delay=10, check=True, skin=2.0)
\end{lstlisting}

In this example, the parameters \texttt{every}, \texttt{delay}, \texttt{check}, and \texttt{skin} control the update frequency and reuse behavior of the neighbor list:
\begin{itemize}
    \item \texttt{every}: interval between neighbor list updates.
    \item \texttt{delay}: buffer before enforcing the next update.
    \item \texttt{check}: whether to validate the reuse condition.
    \item \texttt{skin}: width of the buffer zone (in \AA), allowing small displacements without triggering a rebuild.
\end{itemize}

The implementation is available at:  
\url{https://github.com/CheukHinHoJerry/torch-dftd}.

\subsection{Data Preparation}
\label{sec:sub:data}

The quality of a MLIP is closely tied to the quality of the reference dataset. MLIPs are typically trained via supervised regression on a large set of atomic configurations, each labeled with total energies, atomic forces, and often stress tensors, usually computed from DFT. The same principles apply to fine-tuning U-MLIPs, where a well-designed dataset is crucial for achieving improved accuracy and transferability~\cite{novelli2024fine, focassio2024performance, yu2024systematic, pyzer2025foundation, kulichenko2024data}.

\subsubsection{Empirical Dataset Construction}
\label{sec:sub:sub:empirical}

Empirical dataset construction relies on physically motivated or experience-driven methods. These include using existing public datasets, generating configurations via random perturbations, or sampling configurations from molecular dynamics (MD) or Monte Carlo (MC) simulations. Each method offers distinct advantages and is best suited for particular types of systems or user expertise. In the following, we briefly describe each approach and offer practical recommendations based on our experience.

\paragraph{Using publicly available datasets.}

Depending on observables or quantities of target system, open-access datasets that have been developed for machine-learned interatomic potentials can be readily used for fine-tuning U-MLIPs~\cite{chanussot2021open, bowman2022md17, barroso2024open, aissquare_datasets, nep_data_gitlab, ying2025advances}. A key consideration is format compatibility. For example, the MACE-MP-0 model expects data in the .xyz format, whereas datasets for Deep Potential (DP) models~\cite{wang2018deepmd} are typically provided in .npy format. To address this, we provide a convenient script that converts .npy files into .xyz files:

\begin{lstlisting}[caption=Convert .npy data to .xyz,label=lst:transfer]
python transfer_npy_xyz.py train.npy train.xyz
\end{lstlisting}

Another consideration is dataset size. Public datasets designed for training-from-scratch are often much larger than what is needed for fine-tuning. Since U-MLIPs already encode broad chemical and structural trends, effective fine-tuning can usually be achieved with significantly fewer configurations. This observation will be demonstrated in a case study in Section~\ref{sec:sub:dp_data}.
We note that using publicly available datasets is especially recommended for beginners, as it requires minimal setup and provides immediate access to high-quality reference data.

\paragraph{Random perturbation of structures.}

Another common approach is to apply random displacements to atoms in relaxed structures to sample configurations near local energy minima (local equilibrium). This method is simple and widely used, especially for systems with crystalline structures. If stress data are required, small random deformations of the simulation cell (i.e., strain perturbations) should also be applied. Random perturbations can be generated using ASE’s \texttt{rattle} function, typically with Gaussian-distributed displacements of 0.1–1.0~\AA. Care must be taken to avoid unphysical structures with atoms too close together. This issue can be mitigated by applying a Monte Carlo filtering procedure that enforces a minimum interatomic distance, where configurations with excessively short atomic separations are accepted with low probability~\cite{eriksson2019hiphive}. This method is recommended for users who are familiar with atomistic simulation workflows and have sufficient computational resources. It is particularly well suited for advanced users with prior knowledge of the target system. Notably, when combined with the filtering techniques discussed later, well-designed perturbation strategies can produce compact and physically meaningful datasets that are particularly effective for fine-tuning U-MLIPs~\cite{van2023hyperactive}.

\paragraph{Sampling from MD or MC simulations.}

When no well-defined structural prototype exists, as in amorphous phases or liquids, perturbation methods are not suitable. Instead, ab initio molecular dynamics~\cite{marx2009ab, kresse1993ab} or Monte Carlo simulations provide a more general approach for generating configurations. Snapshots are extracted at regular intervals along the trajectory to build the dataset. Although computationally more demanding, this approach is essential when structural templates are unavailable. To further improve dataset diversity and relevance for fine-tuning, perturbations are often applied to MD-generated structures, and in some cases, MD simulations with enhanced sampling are employed~\cite{bernardi2015enhanced, yang2019enhanced}.

\subsubsection{Uncertainty-Based Dataset Construction}
\label{sec:sub:sub:uncertainty}

Manual dataset design often depends on expert intuition, which may introduce inductive biases, redundant structures, or sampling imbalances. Such issues can undermine data efficiency and hinder the generalizability of trained MLIPs. As a remedy, uncertainty-guided data selection has emerged as a systematic and efficient alternative for both training MLIPs and fine-tuning U-MLIPs~\cite{zhu2023fast, kellner2024uncertainty, edeling2024global, venturi2020bayesian, wen2020uncertainty, tan2023single}.

Uncertainty quantification (UQ) estimates the confidence of a model’s predictions and can be used to identify regions in configuration space where the model is unreliable. Incorporating such estimators enables active learning strategies that prioritize configurations with high predictive uncertainty, thereby improving accuracy while minimizing data redundancy.

Several definitions and methodologies for uncertainty estimation are commonly employed in the context of MLIPs. These approaches can be broadly categorized into frequentist and Bayesian frameworks, as summarized in Table~\ref{tab:uq_methods}. Frequentist techniques include model ensembles~\cite{zhang2019active} and feature-space distance metrics~\cite{smith2018less}, while Bayesian methods rely on posterior inference~\cite{venturi2020bayesian}, dropout-based approximations~\cite{wen2020uncertainty}, or analytical forms available in tractable models such as Gaussian processes and linear regressions~\cite{van2023hyperactive}. In addition, hybrid approaches have been developed to separately quantify aleatoric (data-inherent) and epistemic (model-related) uncertainties~\cite{jiang2024bayesian}. A comprehensive discussion of UQ in MLIPs is beyond the scope of this tutorial. However, we emphasize that this remains an active area of research, and we refer interested readers to recent reviews for further insight~\cite{dai2024uncertainty, grasselli2025uncertainty, wang2025aleatoric}.

\begin{table*}[t]
    \centering
    \caption{Representative uncertainty quantification (UQ) methods used in MLIPs.}
    \label{tab:uq_methods}
    \renewcommand{\arraystretch}{1.2}
    \setlength{\tabcolsep}{5pt}
    \begin{tabular}{l l l}
        \toprule
        \textbf{Method} & \textbf{Key Idea} & \textbf{Category} \\
        \midrule
        Ensemble~\cite{rahaman2021uncertainty} & Use variance across multiple independently trained models & Frequentist \\
        Dropout UQ~\cite{althoff2021uncertainty} & Dropout at inference to sample pseudo-ensembles & Bayesian \\
        Bayesian NN~\cite{olivier2021bayesian} & Infer posterior over weights via VI~\cite{Gal2022Bayesian} or MCMC~\cite{Zhang2021MonteCarlo} & Bayesian \\
        Evidential UQ~\cite{bae2004epistemic} & Predict mean and variance directly as outputs & Evidential \\
        GPR variance~\cite{manfredi2021probabilistic} & Analytical uncertainty from Gaussian processes & Bayesian \\
        Feature distance~\cite{hart2023improvements} & Distance to nearest training point in feature space & Geometric \\
        Aleatoric + Epistemic~\cite{munikoti2023general} & Explicitly model data and model uncertainty & Combined \\
        \bottomrule
    \end{tabular}
\end{table*}

\paragraph{Filtering a candidate set.}
In cases where a large pool of configurations is available (e.g., from long MD simulations or existing datasets), uncertainty-based filtering can be applied to construct an efficient fine-tuning set. Here, an uncertainty estimator (such as an ensemble or a surrogate model with Bayesian regression) is used to assign a confidence value to each configuration. Configurations with higher estimated uncertainty are prioritized, as they are more likely to contribute new information to the model. This strategy has been used successfully in recent works, such as Ref.~\cite{van2023hyperactive}, where a linear ACE model is trained on top of an existing representation to guide uncertainty filtering. We evaluate the data efficiency of fine-tuning by comparing it against uncertainty-based filtering in Section~\ref{sec:sub:Si}.

\paragraph{Active learning strategy.}

Active learning provides a fully iterative framework for dataset construction. Starting from a small initial set of labeled configurations, the MLIP is trained and then used to run MD simulations. Configurations from these simulations are evaluated by a UQ metric, and the most uncertain ones are selected for DFT labeling. The newly labeled data are added to the training set, and the process is repeated until convergence or a stopping criterion is met (e.g., error saturation or maximum data budget). This strategy is particularly effective for exploring diverse chemical spaces or unknown physical regimes.

While uncertainty-driven methods offer significant advantages, they are not universally applicable. Analytical UQ is often limited to linear models or kernel-based methods such as Gaussian process regression. For large, non-linear models like deep neural networks, approximate UQ methods (e.g., ensembles, dropout) may require careful calibration~\cite{kuleshov2018accurate, pernot2023calibration}. Additionally, active learning workflows introduce algorithmic and software complexity, which may be a barrier for new users. 

For novice users, we recommend the following practical guidelines:
(i) Small, chemically narrow systems: Start with a simple ensemble-based UQ (e.g., multiple MACE models trained from different random seeds) and select new configurations with the highest predicted force uncertainty.
(ii) Chemically diverse or poorly explored systems: Consider dropout-based UQ during inference to reduce training cost while still capturing model variance, but apply a post-hoc calibration step such as isotonic regression or temperature scaling.
(iii) When DFT labeling is expensive: Limit the number of configurations per iteration by combining UQ-driven selection with structural diversity filters (e.g., farthest-point sampling in descriptor space~\cite{liu2025farthest}).
These approaches balance exploration efficiency, calibration accuracy, and computational feasibility, enabling active learning workflows that are both effective and accessible to new users.

\subsection{Fine-Tuning}
\label{sec:sub:train}

This section outlines the main steps and considerations for fine-tuning the MACE-MP-0 model. Fine-tuning refers to updating the parameters of a pre-trained U-MLIP on a task-specific dataset, while keeping the overall model architecture fixed. This strategy leverages the broad physical knowledge encoded in the foundation model and enables efficient adaptation to new systems with relatively limited data.

A typical fine-tuning workflow involves constructing the training dataset, (and usually a validation dataset) set with the above mentioned dataset construction methods. Table~\ref{tab:finetune_hyperparams} summarizes representative hyperparameters and their recommended ranges, consistent with the example script provided in Listing~\ref{lst:finetune_mace}.

\begin{table*}[t]
    \centering
    \caption{Representative hyperparameters for fine-tuning MACE-MP-0. The recommended values are representative for the cases considered in this work, and may require adjustment for different systems.}
    \label{tab:finetune_hyperparams}
    \renewcommand{\arraystretch}{1.2}
    \setlength{\tabcolsep}{5pt}
    \begin{tabular}{l l l}
        \toprule
        \textbf{Parameter} & \textbf{Description} & \textbf{Recommended Range} \\
        \midrule
        \texttt{energy\_weight} & Energy loss weight & 1.0 -- 20.0 \\
        \texttt{forces\_weight} & Force loss weight & 1.0 -- 20.0 \\
        \texttt{stress\_weight} & Stress loss weight & 0.0 -- 20.0 \\
        \texttt{lr} & Learning rate & $10^{-4}$ -- $10^{-3}$ \\
        \texttt{batch\_size} & Batch size & 2 -- 20 \\
        \texttt{ema\_decay} & EMA decay rate & 0.98 -- 0.99999 \\      
        \texttt{swa} & Enable stochastic weight averaging & True / False \\
        \texttt{swa\_lr} & Learning rate for SWA phase & $10^{-5}$ -- $ 10^{-4}$ \\
        \bottomrule
    \end{tabular}
\end{table*}

Listing~\ref{lst:finetune_mace} presents a fine-tuning script using the \textsf{run\_train.py} utility provided by the \texttt{MACE} package:

\begin{lstlisting}[language=bash, caption=Fine-tuning template for the MACE-MP-0 model.,label=lst:finetune_mace]
python3 $MACE_DIR/mace/cli/run_train.py \
    --name=$RES_DIR \
    --foundation_model="XXX.model" \
    --model_dir=results/$RES_DIR \
    --log_dir=results/$RES_DIR \
    --checkpoints_dir=results/$RES_DIR \
    --results_dir=results/$RES_DIR \
    --train_file="train.xyz" \
    --valid_file="valid.xyz" \
    --energy_weight=1.0 \
    --forces_weight=10.0 \
    --stress_weight=1.0 \
    --loss="universal" \
    --forces_key=dft_force \
    --energy_key=dft_energy \
    --stress_key=dft_stress \
    --lr=0.0005 \
    --scaling="rms_forces_scaling" \
    --batch_size=4 \
    --max_num_epochs=150 \
    --ema \
    --ema_decay=0.99 \
    --weight_decay=1e-6 \
    --amsgrad \
    --default_dtype="float64" \
    --clip_grad=10 \
    --device=cuda \
    --seed=$SEED \
    --num_samples_pt=500 \
    --swa_energy_weight=100.0 \
    --swa_forces_weight=10.0 \
    --swa_stress_weight=1.0 \
    --swa \
    --swa_lr=5e-4 \
    --E0s="{3:-0.29745415, 15:-1.88719893, 16:-1.08090657, 32:-0.77926738}" \
\end{lstlisting}

From our experience, a medium-sized model ($L=1$) provides the most balanced trade-off between accuracy and computational efficiency. While the small model ($L=0$) is computationally inexpensive but lacks expressivity, the large model ($L=2$) increases accuracy at the cost of significantly reduced efficiency. The following recommendations are in general helpful on stability and accuracy of the resulting model from fine-tuning:

\begin{itemize}
    \item Isolated atom energies (\texttt{E0s}): It is very important to compute your own \texttt{E0s}, which specify the isolated atom energy as a dictionary in your DFT calculation rather by estimating it from least square regression with \texttt{E0s="average"}. When computing your \texttt{E0s}, use spin polarized calculations. You may also use \texttt{E0s=”foundation”} to use the same \texttt{E0s} when using exactly the same DFT settings and software as Materials Project.
    \item Learning rate (\texttt{lr}): If the training or validation loss stagnates or diverges, consider modifying the learning rate from recommended range for more stable convergence.
    \item Loss weights: Depending on the target properties, modify the ratio between energy, forces and stress weight in fine-tuning within the recommended range.
    \item Cutoff radius (\texttt{r\_max}): Do not modify the pre-defined cutoff radius that the foundation model used. 
    \item Batch size: A larger batch size generally improves training stability but requires more GPU memory. Batch size larger than 20 is not recommended since a larger batch size, in general, tend to deteriorate the generalizibility of machine learning models.
    \item EMA and SWA: Enable exponential moving average (\texttt{ema}) and stochastic weight averaging (\texttt{swa}) for smoother convergence and better generalization. If training dynamics appear unstable, set \texttt{swa} to 0.99999. The model with best validation error from either pre-\texttt{swa} or post-\texttt{swa} based (i.e. \texttt{stageone} and \texttt{stagetwo} in Section~\ref{sec:sub:validation}) on validation loss are outputted to be further verified on other properties.
    \item Number of samples in the pretrained head (\texttt{num\_samples\_pt}): This parameter controls how many samples from the pretrained model’s output distribution are retained for the fine-tuning stage. A larger value preserves a richer set of features and variance from the pretrained head, which can improve MD stability. In other words, it acts as a form of “knowledge retention” that mitigates catastrophic forgetting of the pretrained model. However, there is a trade-off: if it is too large, the fine-tuned model remains overly biased towards the pretrained distribution, which may limit its ability to fit task-specific data and thus cap the achievable accuracy.
\end{itemize}

We recommend monitoring convergence behavior (cf.~Fig.~\ref{fig:loss}) and validation accuracy throughout the fine-tuning process. Evaluating model performance on unseen configurations is crucial and will be discussed in the next subsection.V

\subsection{Model Validation and Evaluation}
\label{sec:sub:validation}

After completing the fine-tuning process, several output files are generated by the training script. Among them, the most critical is the log file, which contains all essential information needed to reproduce the results. This includes the selected hyperparameters, loss trajectories over all training epochs, and final test errors (typically reported in terms of RMSE). Toward the end of the log, a summary table presents test errors for energies, forces, and—if available—stresses.

Additionally, the script automatically generates diagnostic plots that visualize the training and validation loss curves, as well as scatter plots comparing predicted and reference energies and forces. These visualizations provide a qualitative assessment of model performance. In particular, smooth loss convergence and tightly clustered scatter plots are indicative of successful fine-tuning.

\begin{figure*}
\centering
\includegraphics[width=17 cm]{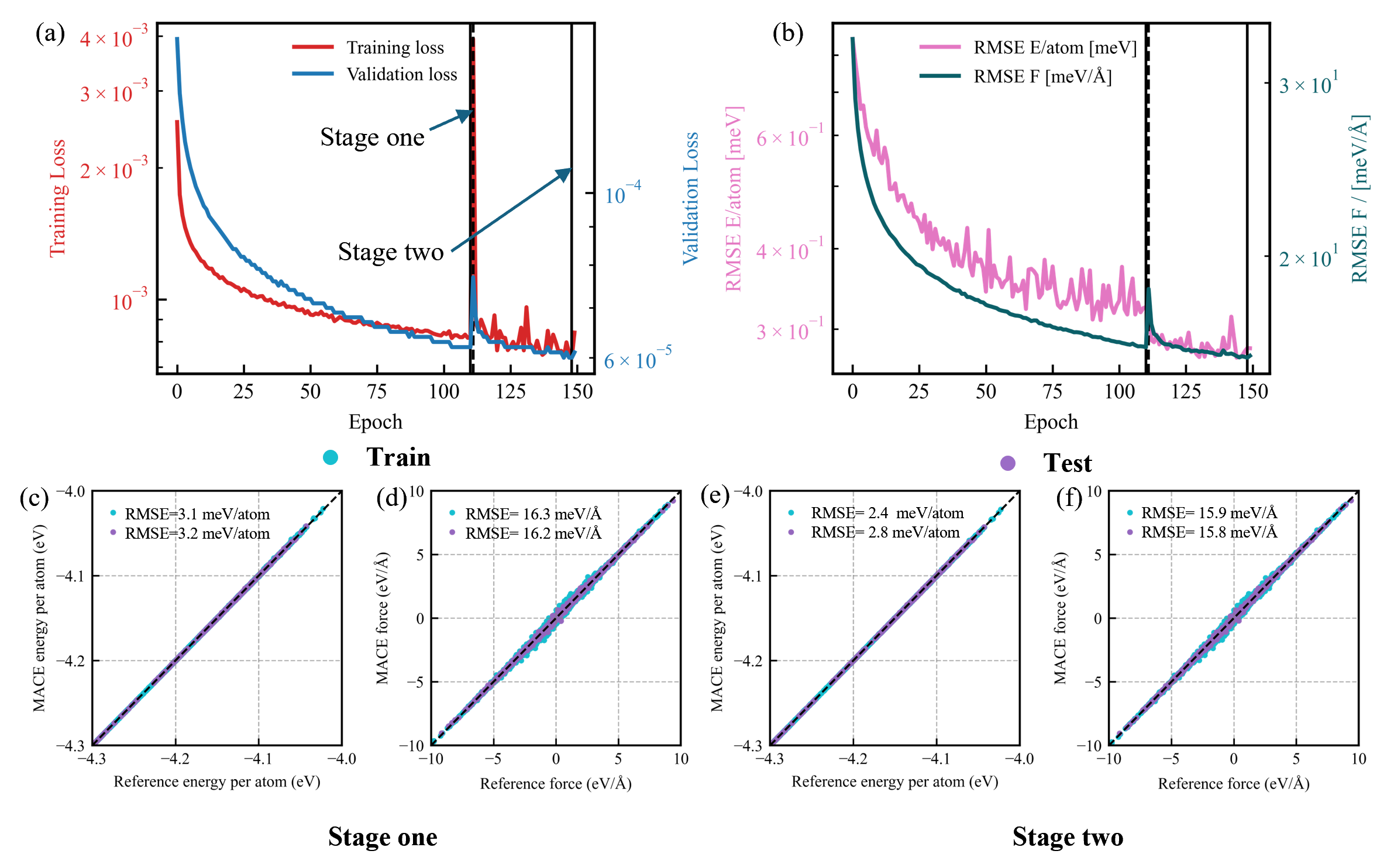}
\caption{Training and validation loss curves, and fitting accuracy for energy and force predictions. The left panel corresponds to the initial training stage (before stochastic weight averaging), while the right panel shows results after SWA.}
\label{fig:loss}
\end{figure*}

As an illustration, we refer to the example in Section~\ref{sec:sub:dp_data}, where such diagnostic plots are used to evaluate the performance of a fine-tuned model. Figure~\ref{fig:loss} displays the evolution of training and validation losses, along with the final energy and force scatter plots.

The output directory also contains several key model files:
\begin{itemize}
    \item \texttt{MACE-FT.model}: the complete fine-tuned model used for inference;
    \item \texttt{MACE-FT\_compiled.model}: a compiled version optimized for deployment;
    \item \texttt{MACE-FT\_stageone.model} and \texttt{MACE-FT\_stagetwo.model}: intermediate models from different training stages (e.g., before and after stochastic weight averaging).
\end{itemize}

These models can be directly used in downstream atomistic simulations and analysis workflows to evaluate a variety of important material properties. These include structural and mechanical properties such as bulk modulus and defect formation and migration energies, thermodynamic properties such as density, pressure, and phonon spectra, as well as dynamical properties such as radial distribution functions and diffusion coefficients obtained from molecular dynamics simulations.  

\paragraph{Example: Structural Relaxation with ASE.}

Once fine-tuned, the model can be seamlessly integrated into the Atomic Simulation Environment (ASE) for structural optimization and property evaluation. The example below demonstrates how to perform structural relaxation on a given atomic configuration using the \texttt{MACECalculator} within ASE:

\begin{lstlisting}[caption=Structural relaxation using a fine-tuned MACE model,label=lst:ace_optima]
from mace.calculators import MACECalculator
ase_atoms.calc = MACECalculator("MACE-FT.model")

from ase.optimize.precon import PreconLBFGS
optimizer = PreconLBFGS(ase_atoms, variable_cell=True)
optimizer.run(fmax=1e-5)
\end{lstlisting}

This procedure minimizes both atomic positions and cell parameters using the preconditioned L-BFGS algorithm. The same workflow can be extended to defect formation energy calculations, nudged elastic band (NEB) transition state searches, or molecular dynamics (MD) simulations.

\paragraph{Example: Using the Model in LAMMPS.}

To use the fine-tuned model in LAMMPS, it must first be converted to a LAMMPS-compatible format using the provided utility script:

\begin{lstlisting}[caption=Exporting the model for LAMMPS,label=lst:lammps]
python <mace_repo_dir>/mace/cli/create_lammps_model.py MACE-FT.model
\end{lstlisting}


Executing the above command generates a LAMMPS-compatible model file named \texttt{MACE-FT.model-lammps.pt}, which can be directly used in LAMMPS simulations via the following commands:
\begin{lstlisting}[caption=Using the fine-tuned MACE potential in LAMMPS,label=lst:paircoeff]
pair_style    mace no_domain_decomposition
pair_coeff    * *  MACE-FT.model-lammps.pt elmemt1 elment2 elment3
\end{lstlisting}

Here, \texttt{pair\_style} specifies the MACE interaction style without domain decomposition, and \texttt{pair\_coeff} is used to load the compiled model file along with the corresponding atomic species. In our molecular dynamics simulations, we adopted the \texttt{metal} unit style and applied a canonical (NVT) ensemble to maintain the temperature at 298~K. 
%


We refer readers to our third case study (Section~\ref{sec:sub:gra_water}) for a complete example of running MD simulations in LAMMPS using a fine-tuned model. All input files and scripts are available in our public tutorial repository:  
\url{https://github.com/John2021-hub/mace-ft-tutorial.git}.

\section{EXAMPLES AND APPLICATIONS}
\label{sec:examples}

The fine-tuned models integrate seamlessly with widely used atomistic simulation platforms such as ASE, LAMMPS and RBMD (see Appendix~\ref{sec:apd:codes}), enabling efficient molecular dynamics and other atomistic simulations. In this section, we illustrate the effectiveness of fine-tuning using representative case studies on solid-state electrolytes, metallic stacking faults, and interfacial phenomena in low-dimensional systems. The pretrained foundation models are publicly available at \url{https://github.com/ACEsuit/mace-mp/}. In addition, we provide fine-tuning scripts, example outputs, and best practices in our tutorial repository: 
\url{https://github.com/John2021-hub/mace-ft-tutorial.git}.

We performed fine-tuning and molecular dynamics simulations on a single NVIDIA A800 GPU with 80 GB of HBM memory. The software environment was configured with CUDA 11.7 and the NVIDIA HPC SDK (nvhpc) 20.9. Unless otherwise specified, all training runs were carried out under this single-GPU setup.

We present four representative examples to illustrate the fine-tuning of U-MLIPs across different materials contexts. Solid-state electrolytes are chosen to assess improvements in force prediction essential for accurate ion-transport simulations. Stacking fault defects in metals are included to evaluate generalization to defect configurations and the accurate prediction of stacking fault energies. Silicon with a diverse and complex dataset is used to evaluate out-of-distribution generalization. Solid–liquid interfacial systems are examined to test the model’s ability to capture interfacial energetics critical to phase transitions and growth processes.

\subsection{Li\textsubscript{10}GeP\textsubscript{2}S\textsubscript{12}}
\label{sec:sub:dp_data}

In this example, we demonstrate the fine-tuning procedure on a solid-state electrolyte system, lithium germanium phosphorous sulfide (Li\textsubscript{10}GeP\textsubscript{2}S\textsubscript{12}, abbreviated as LGPS), using a dataset published on DPAISquare~\cite{aissquare_datasets}. LGPS is a prototypical member of the LGPS-type thiophosphate family, and has attracted significant attention due to its high lithium-ion conductivity, making it a promising candidate for next-generation all-solid-state batteries. Modeling such materials requires accurate interatomic potentials capable of describing both local structural features and ionic transport mechanisms, making LGPS an ideal testbed for evaluating the transferability and fine-tuning of U-MLIPs.

\begin{figure}
\centering
\includegraphics[height=5.5cm]{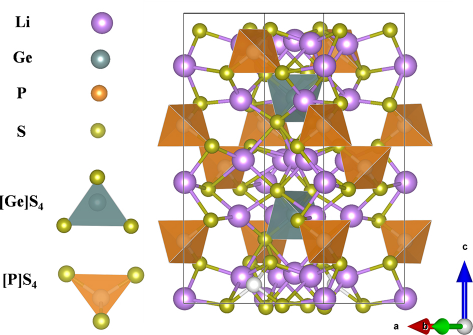}
\caption{Atomic structure of Li\textsubscript{10}GeP\textsubscript{2}S\textsubscript{12}, a solid-state electrolyte with high lithium-ion conductivity.}
\label{fig:structure_LiGePS}
\end{figure}

The LGPS dataset published on DPAISquare~\cite{aissquare_datasets} can be converted into the .xyz format supported by MACE, making it directly usable in the fine-tuning workflow (cf.~Listing~3). Figure~\ref{fig:structure_LiGePS} shows the atomic structure of a typical configuration from this dataset.

Table~\ref{tab:fittingaccuracy} summarizes the test accuracy of multiple MACE foundation models after fine-tuning. All models achieve comparable accuracy in terms of root-mean-square error (RMSE) with respect to DFT reference values. The two rows for each model correspond to results obtained before and after the application of stochastic weight averaging (SWA), where the first epoch indicates the SWA start step, and the second corresponds to the final averaged model. The consistency of test accuracy before and after SWA highlights the stability and robustness of the fine-tuning procedure. Also, the key point is that for all U-MLIPs, with the same hyperparameters, we can all get comparable accuracy, showing that the robustness of MACE-based foundation models and its fine-tuning.

\begin{table*}[t]
\centering
\caption{\label{tab:fittingaccuracy} 
Test accuracy of different fine-tuned MACE foundation models, reported as RMSE of energy (E) and force (F), relative to DFT values. “Relative F” denotes the force RMSE normalized by the standard deviation of DFT forces. Each model is evaluated before and after SWA, with the epoch number indicating the training step. Fine-tuning consistently achieves good accuracy regardless of which foundation model is used. } 
{\renewcommand{\arraystretch}{1.3}
\begin{ruledtabular}
\begin{tabular}{lccc}
Model (epoch) & RMSE E (meV/atom) & RMSE F (meV/\AA) & Relative F (\%) \\
\hline
MACE-MP-0b3 (110)  & 0.32 & 16.15 & 1.92 \\
MACE-MP-0b3 (148)  & 0.28 & 15.84 & 1.87 \\
MACE-MPA-0 (110)  & 0.30 & 15.07 & 1.79 \\
MACE-MPA-0 (146)  & 0.27 & 14.88 & 1.77 \\
MACE-OMAT-0 (110) & 0.35 & 15.46 & 1.84 \\
MACE-OMAT-0 (149) & 0.27 & 15.32 & 1.82 \\
\end{tabular}
\end{ruledtabular}
}
\end{table*}

\begin{figure*}
\centering
\includegraphics[height=6cm]{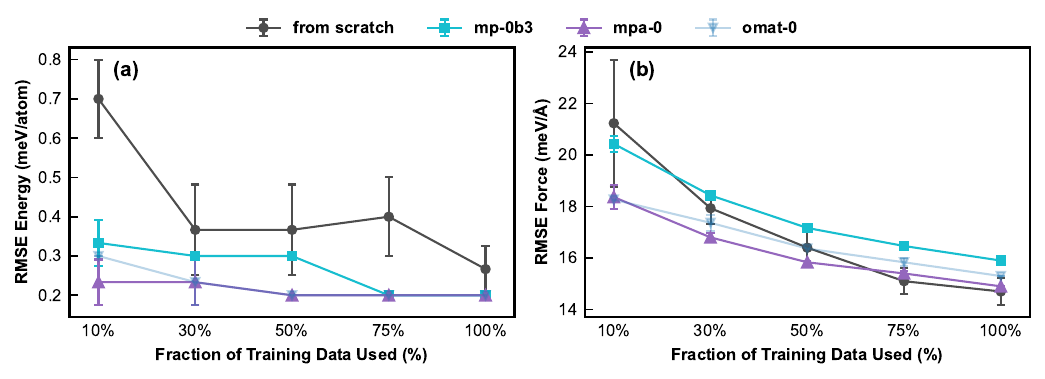}
\caption{Comparison between fine-tuning and training-from-scratch in terms of data efficiency: (a) RMSE in energy (meV/atom) and (b) RMSE in forces (meV/\AA). Error bars represent one standard deviation, computed from three independent runs, with markers showing the corresponding mean values.}
\label{fig:ft_efficiency}
\end{figure*}

Figure~\ref{fig:ft_efficiency} compares the test accuracy of models fine-tuned from different MACE foundation models with that of training-from-scratch, evaluated at different fractions of the training dataset (10\%, 30\%, 50\%, 75\% and 100\%). Figure~\ref{fig:ft_efficiency}(a) shows the RMSE in energy, while Figure~\ref{fig:ft_efficiency}(b) shows the RMSE in forces. Across all dataset sizes, fine-tuned models (MP-0b3, MPA-0, OMAT-0) consistently achieve lower errors than training-from-scratch, particularly at lower data fractions, demonstrating the data efficiency and accuracy benefits of fine-tuning. The error bars remain small across experiments, indicating robustness against training noise. Overall, these results highlight the strength of MACE foundation models as effective starting points for fine-tuning, enabling improved performance even with limited training data.

\subsection{Silicon}
\label{sec:sub:Si}

We use silicon studied in \cite{bartok2018machine} as a representative system to evaluate the out-of-distribution generalization of fine-tuning, since the potential risk of overfitting has attracted considerable attention in the fine-tuning of foundation models. The training set consists of standard bulk and defect configurations, while the test set includes grain boundaries, di-interstitials, stacking fault paths, and an amorphous configuration. Unlike a random split, the test set contains configurations entirely distinct from the training data, providing a stringent benchmark for assessing extrapolative performance.

The previous example highlighted the data efficiency of fine-tuning with limited training data. In this case, we focus on ensemble-uncertainty-based active learning, where candidate configurations are filtered according to prediction uncertainty (cf.~Section~\ref{sec:sub:sub:uncertainty}) following the procedure in~\cite{van2023hyperactive}. This reduces the training set to only 5\% of the original data. We then train a MACE model on this reduced dataset and compare its accuracy with that obtained from direct fine-tuning.

\begin{figure*}
\centering
\includegraphics[width=16.5cm]{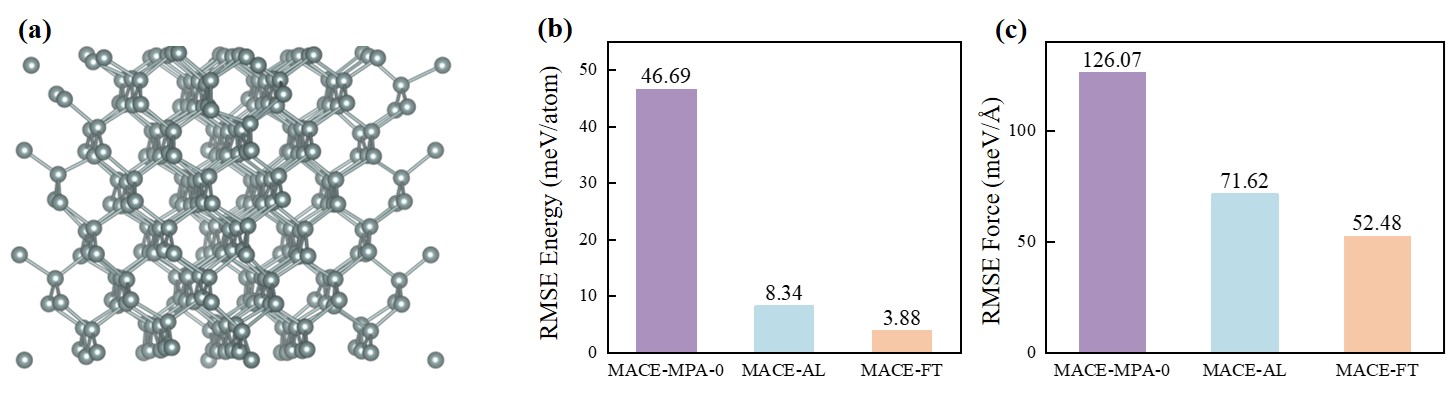}
\caption{Comparison of a foundation MACE model (MACE-MPA-0) and its fine-tuned counterpart (MACE-FT) on the silicon out-of-distribution test set. (a) Atomic structure; (b) energy RMSE per atom (meV/atom); (c) force RMSE (meV/Å). In both cases, MACE-FT achieves significantly lower errors than MACE-MPA-0, demonstrating improved accuracy and generalization. Moreover, MACE-FT also outperforms models trained from uncertainty-filtered datasets (MACE-AL).
}
\label{fig:Si_RMSE}
\end{figure*}

Figure~\ref{fig:Si_RMSE} compares the performance of the foundation MACE model (MACE-MPA-0) with its fine-tuned counterpart (MACE-FT): (a) energy RMSE per atom (meV/atom), and (b) force RMSE (meV/Å). In both metrics, MACE-FT achieves substantially lower errors than MACE-MPA-0, indicating improved agreement with the reference data even for the out-of-distribution test set. This demonstrates that fine-tuning does not suffer from generalization issues in foundation models. A second conclusion is drawn from the comparison with models trained on uncertainty-filtered datasets (labeled “MACE-AL” in Figure~\ref{fig:Si_RMSE}). Although the filtered set contains a similar number of samples, fine-tuning still yields superior accuracy. This highlights that fine-tuning not only ensures generalization but also achieves better data efficiency than training from scratch on a reduced dataset obtained through UQ-based active learning.

\subsection{Molybdenum}
\label{sec:sub:mo}

In this example, we evaluate the capability of fine-tuned U-MLIPs to accurately describe mechanical properties associated with crystalline defects in metallic systems. Molybdenum (Mo), a body-centered cubic (BCC) transition metal, is chosen as a representative case due to its well-characterized elastic properties and defect energetics, including vacancy formation and migration~\cite{naghdi2024neural}. 

\begin{figure}
\centering
\includegraphics[height=6.5cm]{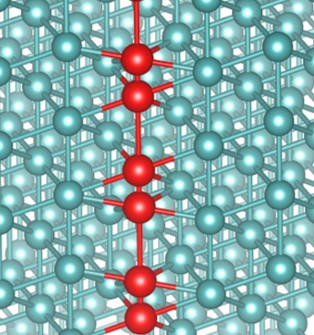}
\caption{Atomic structure of BCC Mo containing a dislocation and the associated local lattice distortion. The defect core is highlighted in red.}
\label{fig:structure_Mo}
\end{figure}

Crystalline defects play a key role in determining the plasticity, strength, and diffusion behavior of metals. However, accurately capturing their highly localized and stress-sensitive nature remains a challenge for many interatomic potentials. This example is therefore designed to assess whether fine-tuning improves the quantitative prediction of defect-related energies and transition states in Mo. The atomic structure of bulk Mo is shown in Figure~\ref{fig:structure_Mo}. The dataset used for fine-tuning is taken from Ref.~\cite{naghdi2024neural}.

{\it Bulk validation} ---
Table~\ref{tab:mo} compares the elastic constants ($C_{11}$, $C_{12}$, $C_{44}$), bulk modulus ($B$), and Poisson ratio ($\nu$) of molybdenum predicted by different interatomic potentials (EAM, GAP, MACE-MP-0b3, and fine-tuned MACE-MP-0) against DFT and experimental reference values. The numbers in parentheses indicate the percentage error relative to experiment.
The results show that fine-tuning the MACE-MP-0 model significantly improves its accuracy across all properties, reducing the error from 45.91\% to 2.58\% for $C_{11}$, from 56.88\% to 14.77\% for $C_{44}$, and from 48.27\% to 3.45\% for $\nu$. While the original MACE-MP-0b3 underestimates elastic constants and bulk modulus, fine-tuning closes the gap to both DFT and experimental values, achieving accuracy comparable to or better than EAM and GAP in most properties. These results highlight the effectiveness of fine-tuning in improving the mechanical property predictions of U-MLIPs.

\begin{table*}
\caption{\label{tab:mo} Elastic constants $C_{ij}$, bulk modulus $B$, and Poisson ratio $\nu$, as obtained with the GAP, tabGAP, EAM/FS, and the NNIP potentials, compared to DFT calculations performed in this work and experimental data. Values in parentheses report the magnitude of the percentage error with respect to the experimental value.} 
{\renewcommand{\arraystretch}{1.3}
\begin{ruledtabular}
\begin{tabular}{lcccccc}
 & EAM~\cite{smirnova2013ternary} & GAP~\cite{byggmastar2020gaussian} & MACE-MP-0b3 & Fine-tuning & DFT~\cite{naghdi2024neural} & Experiment~\cite{crc2017handbook} \\
\hline
$C_{11}$ (GPa) & 465 (0.22\%) & 478 (3.02\%) & 251 (45.91\%) & 452 (2.58\%) & 459 & 464 \\
$C_{12}$ (GPa) & 161 (1.26\%) & 166 (4.40\%) & 189 (18.87\%) & 174 (9.43\%) & 162 & 159 \\
$C_{44}$ (GPa) & 109 (0\%) & 108 (0.92\%) & 47 (56.88\%) & 82 (14.77\%) & 97 & 109 \\
$B$ (GPa) & 263 (5.20\%) & 270 (8.00\%) & 210 (16.00\%) & 267 (6.80\%) & 262 & 250 \\
$\nu$ & 0.26 (10.34\%) & 0.26 (10.34\%) & 0.43 (48.27\%) & 0.28 (3.45\%) & 0.30 & 0.29 
\end{tabular}
\end{ruledtabular}
}
\end{table*}

\begin{figure*}
\centering
\includegraphics[height=7cm]{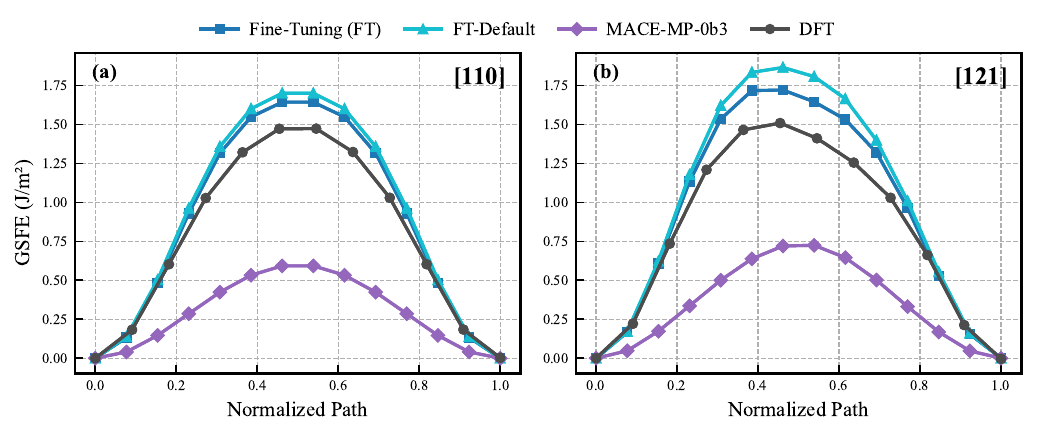}
\caption{GSFE profiles of Mo along the $\langle110\rangle$ (panel (a)) and $\langle112\rangle$ (panel (b)) directions. Two fine-tuned models are compared: FT-Default, which uses the default loss weights, and FT, which employs a force loss weight of 100.0. }
\label{fig:Mo_gsfe}
\end{figure*}

{\it Generalized stacking fault energy} ---
Figure~\ref{fig:Mo_gsfe} shows the generalized stacking fault energy (GSFE) profiles of Mo along the $\langle110\rangle$ (panel (a)) and $\langle112\rangle$ (panel (b)) directions. Table~\ref{tab:gsfe_error} provides the corresponding GSFE barrier errors relative to DFT. The results compare the MACE foundation model, two fine-tuned models (FT-Default, which employs the default loss weights, and FT, which uses an increased force loss weight of 100.0), and the DFT reference data. The foundation model clearly underestimates the GSFE in both directions, leading to lower energy barriers. After fine-tuning, the predictions become much closer to the DFT values, especially near the peak and along the entire fault path. Between the two fine-tuned models, FT performs slightly better than FT-Default, suggesting that adjusting the force weighting can further improve accuracy.

\begin{table*}[t]
\centering
\caption{\label{tab:gsfe_error} GSFE barrier errors relative to DFT are reported for the $\langle110\rangle$ and $\langle121\rangle$ slip directions, comparing three interatomic potential models: Fine-Tuning, FT-Default, and MACE-MP-0b3.} 
{\renewcommand{\arraystretch}{1.3}
\begin{ruledtabular}
\begin{tabular}{lccc}
Slip Directions & Fine-Tuning & FT-Default & MACE-MP-0b3 \\
\hline
$\langle110\rangle$  & \textbf{0.17} & 0.23 & 0.88 \\
$\langle121\rangle$  & \textbf{0.21} & 0.36 & 0.78 \\
\end{tabular}
\end{ruledtabular}
}
\end{table*}

\subsection{Graphene–Water Interface}
\label{sec:sub:gra_water}

\begin{figure}
\centering
\includegraphics[height=7cm]{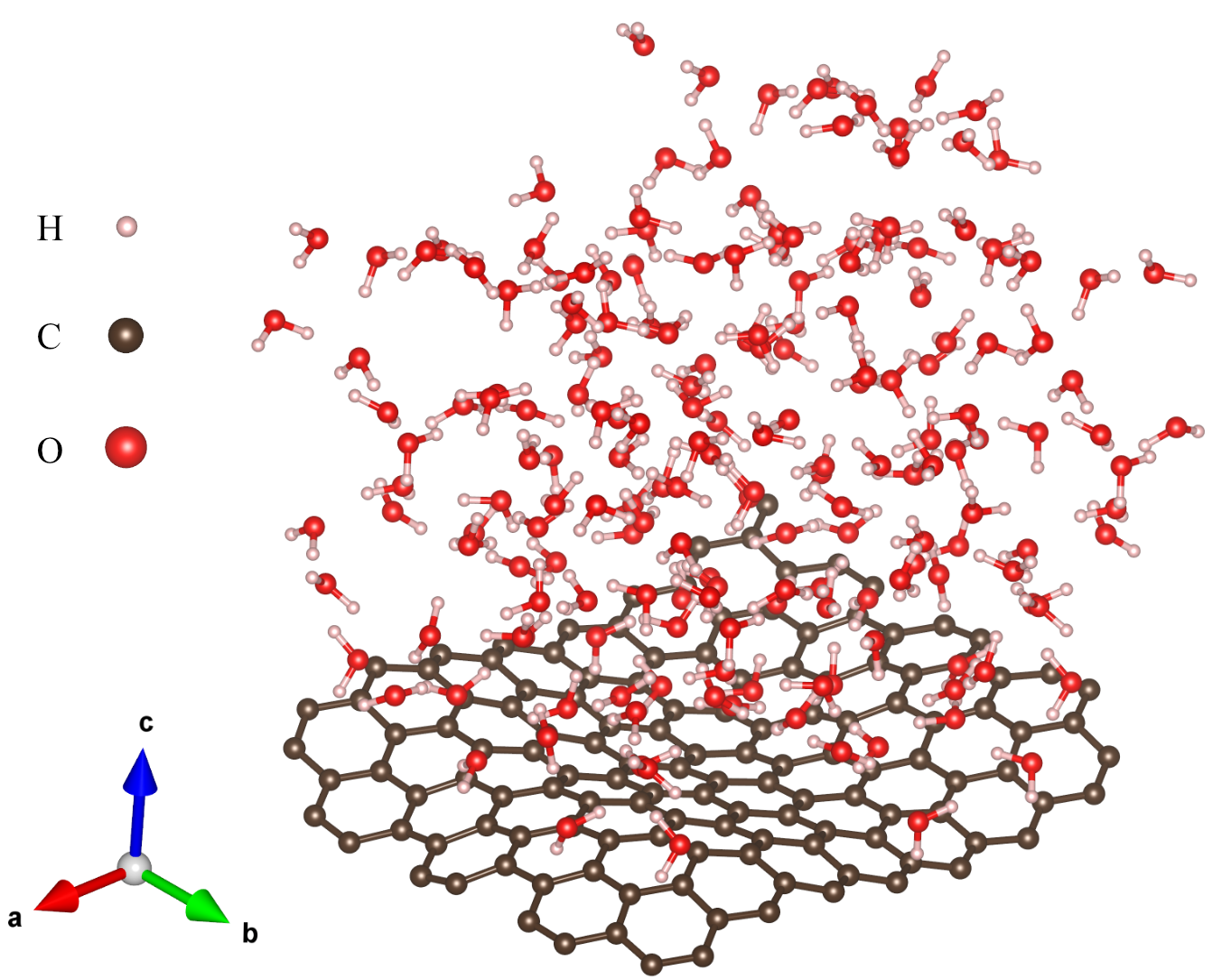}
\caption{Atomic structure of the graphene–water interface used in this study.}
\label{fig:structure_GraWater}
\end{figure}

Graphene–water interfaces are prototypical solid–liquid systems with wide-ranging applications in sensing, electrochemistry, and nanofluidics. Capturing the interfacial structure and dynamics of such systems presents a significant challenge for interatomic potentials, especially in describing long-range interactions and subtle surface effects. This example illustrates the capability of fine-tuned U-MLIPs to model solid–liquid interfaces accurately, using a representative graphene–water system. The atomic structure of the graphene–water interface is illustrated in Figure~\ref{fig:structure_GraWater}. The system contained over 372 atoms and was simulated for 200~ps with a 1~fs timestep.

The graphene–water dataset used to fine-tune the U-MLIP was generated by pre-equilibrating a periodic graphene–water interface with classical molecular dynamics (LAMMPS) \cite{plimptonFastParallelAlgorithms1995,thompsonLAMMPSFlexibleSimulation2022}. Graphene was modeled with a Tersoff bond-order potential \cite{tersoffEmpiricalInteratomicPotential1988,tersoffModelingSolidstateChemistry1989}, and water with a rigid three-site model constrained by SHAKE; short-range interactions were described by Lennard–Jones terms \cite{werderWaterCarbonInteractionUse2003}, and long-range electrostatics by Ewald summation under periodic boundary conditions. Canonical (NVT) sampling furnished equilibrated configurations, from which temporally uncorrelated snapshots were randomly selected for ab initio labeling. Single-point DFT labels were obtained in VASP \cite{kresseEfficientIterativeSchemes1996} using the PAW \cite{blochlImprovedTetrahedronMethod1994} formalism and the PBE exchange–correlation functional \cite{hammerImprovedAdsorptionEnergetics1999} augmented with D3 dispersion \cite{grimmeEffectDampingFunction2011} to capture graphene–water physisorption. cutoff of 400\,eV was chosen for the plane‑wave basis, and electronic self‑consistency was achieved when the total‐energy change fell below $1\times10^{-6}$\,eV.  
Geometry optimisations were considered converged once the residual forces on all atoms were smaller than $0.02$\,eV\,\AA$^{-1}$.

\begin{figure*}
\centering
\includegraphics[height=6.5 cm]{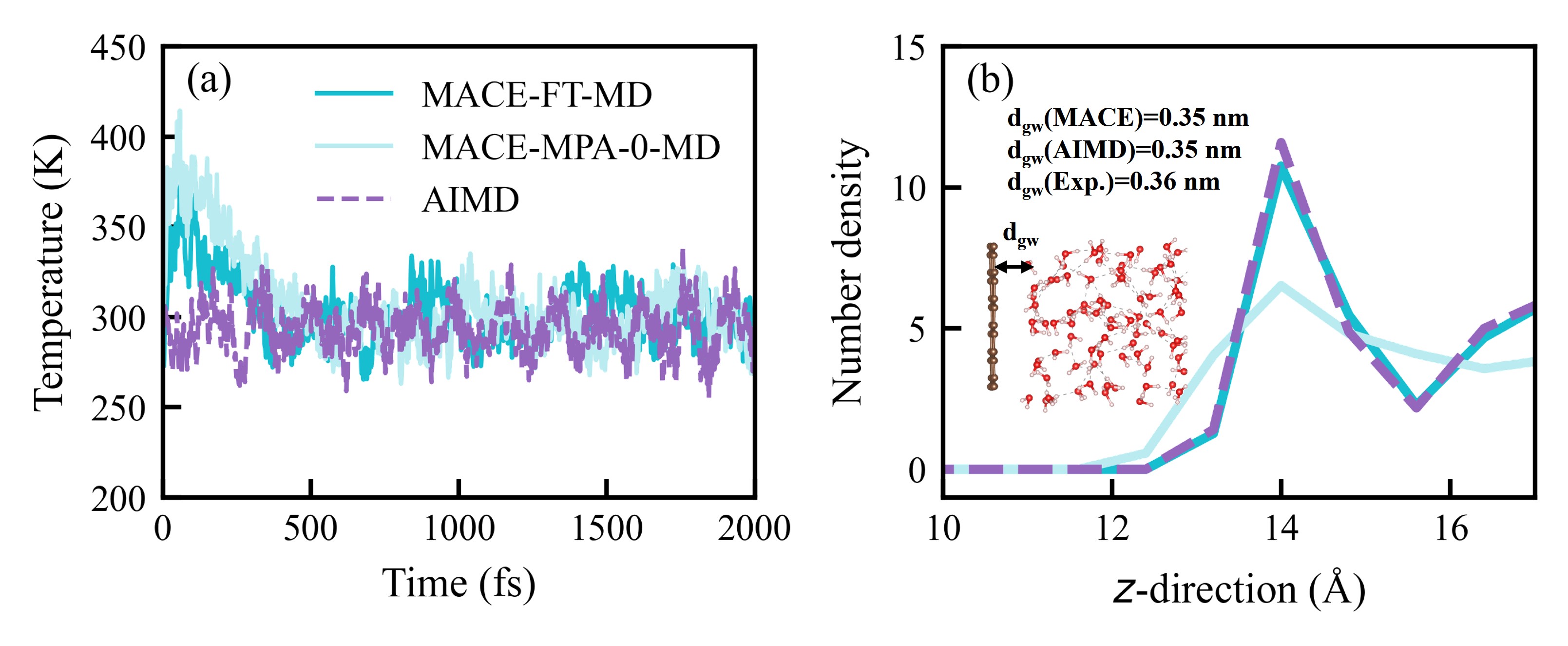}
\caption{Comparison of graphene–water interface simulations using MACE-FT-MD, MACE-MPA-0-MD and AIMD. (a) Temperature evolution over time. (b) Oxygen number-density profile along the z-axis; The first peak corresponds to the graphene–water distance, ${\rm d}_{\rm gw} = 0.35~{\rm nm}$, which is in good agreement with the experimental value of 0.36 nm~\cite{uhligAtomicscaleMappingHydrophobic2019}. The inset illustrates the definition of ${\rm d}_{\rm gw}$. To make a comparison, we also plot the MD trajectory for MACE-MPA-0 model without fine-tuning.}
\label{fig:MD}
\end{figure*}

Figure~\ref{fig:MD} compares molecular dynamics simulations of a graphene–water interface performed using LAMMPS-MACE with the fine-tuned model (MACE-FT-MD) and ab initio molecular dynamics (AIMD). Figure~\ref{fig:MD}(a) shows the temperature evolution, with both simulations maintaining comparable temperature fluctuations around 300 K. Figure~\ref{fig:MD}(b) presents the oxygen number-density profile along the $z$-axis. The first peak corresponds to the graphene–water interfacial distance, ${\rm d}{\rm gw} = 0.35$ nm, which is in good agreement with the experimental value of 0.36 nm~\cite{uhligAtomicscaleMappingHydrophobic2019}. Due to the hydrophobic nature of graphene, a depletion layer forms at the interface, and ${\rm d}{\rm gw}$ is defined to quantify this interfacial gap. This distance is a critical physical quantity for evaluating the accuracy of the interaction potentials and demonstrates the effectiveness of the fine-tuned model.

Table~\ref{tab:scaling} reports the computational cost (wall-clock time in seconds) for 10,000-step molecular dynamics simulations performed with and without the {\tt cuEquivalence} kernel, across systems of varying sizes (128 to 2900 atoms). For each configuration, simulations using cuEquivalence (``With") show consistently lower elapsed times compared to runs without it (``Without"), especially for larger systems. For instance, at 2900 atoms, {\tt cuEquivalence} reduces the runtime from 5084 seconds to 2619 seconds. The scaling factor improves from 0.76 (without {\tt cuEquivalence}) to 0.41 (with {\tt cuEquivalence}), indicating better parallel efficiency with the kernel enabled. These results suggest that the new kernel can serve as an efficient building block for large-scale and long-timescale molecular dynamics simulations.

\begin{table}
\caption{\label{tab:scaling} Computational cost (wall-clock time in seconds) for 10000-step molecular dynamics simulations with and without the use of the \textsf{cuEquivalence} kernel. The first column indicates the total number of atoms in the system, while the remaining columns report the elapsed time for each configuration. “With” denotes simulations utilizing the \textsf{cuEquivalence} kernel, and “Without” corresponds to runs without it.} 
{\renewcommand{\arraystretch}{1.3}
\begin{ruledtabular}
\begin{tabular}{lccccc}
Atoms & 128 & 372 & 1038 & 2900 & Scaling \\
\hline
With (s) & 710 & 800 & 1126 & 2619 & \textbf{0.41} \\
Without (s) & 658 & 1008 & 2031 & 5084 & 0.76 
\end{tabular}
\end{ruledtabular}
}
\end{table}

\subsection{Molecule/Oxide Interface}
\label{sec:sub:au-mgo}

As a representative case of molecule–oxide interactions, we consider a gold dimer (Au$_2$) adsorbed on the MgO(001) surface with aluminum doping, following~\cite{kim2024learning}. The dataset is also from the authors of ~\cite{kim2024learning}. This system has been widely employed to benchmark long-range effects such as charge transfer and polarization at oxide surfaces. It thus provides a stringent test of whether fine-tuned foundation models can accurately capture long-range physics beyond the reach of conventional MLIPs.

\begin{figure*}
\centering
\includegraphics[width=16.5cm]{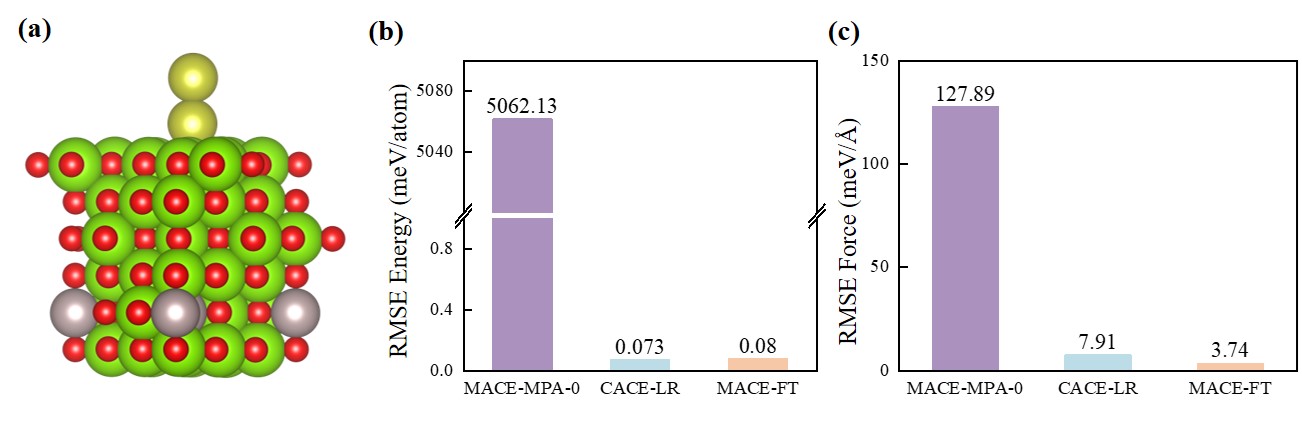}
\caption{Performance comparison between a foundation MACE model (MACE-MPA-0) and its fine-tuned counterpart (MACE-FT) on the doped Au-MgO surface. (a) Atomic structure; (b) energy RMSE per atom (meV/atom); (c) force RMSE (meV/Å). MACE-FT achieves consistently lower errors than MACE-MPA-0 and reaches accuracy comparable to CACE-LR~\cite{kim2024learning}, a state-of-the-art method with explicit long-range interactions. 
}
\label{fig:AuMgO_RMSE}
\end{figure*}

Figure~\ref{fig:AuMgO_RMSE} compares the performance of the foundation MACE model (MACE-MPA-0) with its fine-tuned counterpart (MACE-FT): (a) energy RMSE per atom (meV/atom), and (b) force RMSE (meV/Å). In both metrics, MACE-FT achieves substantially lower errors than MACE-MPA-0, demonstrating improved agreement with the reference data (lower is better). The relatively large error of MACE-MPA-0 arises because its reported value corresponds to the step-0 output during training, reflecting a mismatch between the DFT settings used in this dataset and those employed in the original MPA training. Notably, even without explicitly modeling long-range interactions, the fine-tuned foundation model outperforms recent state-of-the-art methods reported in~\cite{kim2024learning}. The CACE-LR model, which incorporates a latent Ewald summation to capture long-range effects, already delivers strong accuracy; yet our fine-tuned model achieves comparable or even superior performance. This suggests that fine-tuning can implicitly adapt to capture aspects of long-range physics, highlighting the potential of foundation models as a flexible and powerful basis for tackling complex interfacial systems.

\subsection{Solid–Solid Interface}
\label{sec:sub:interfaces}

\begin{figure}
\centering
\includegraphics[width=7.5cm]{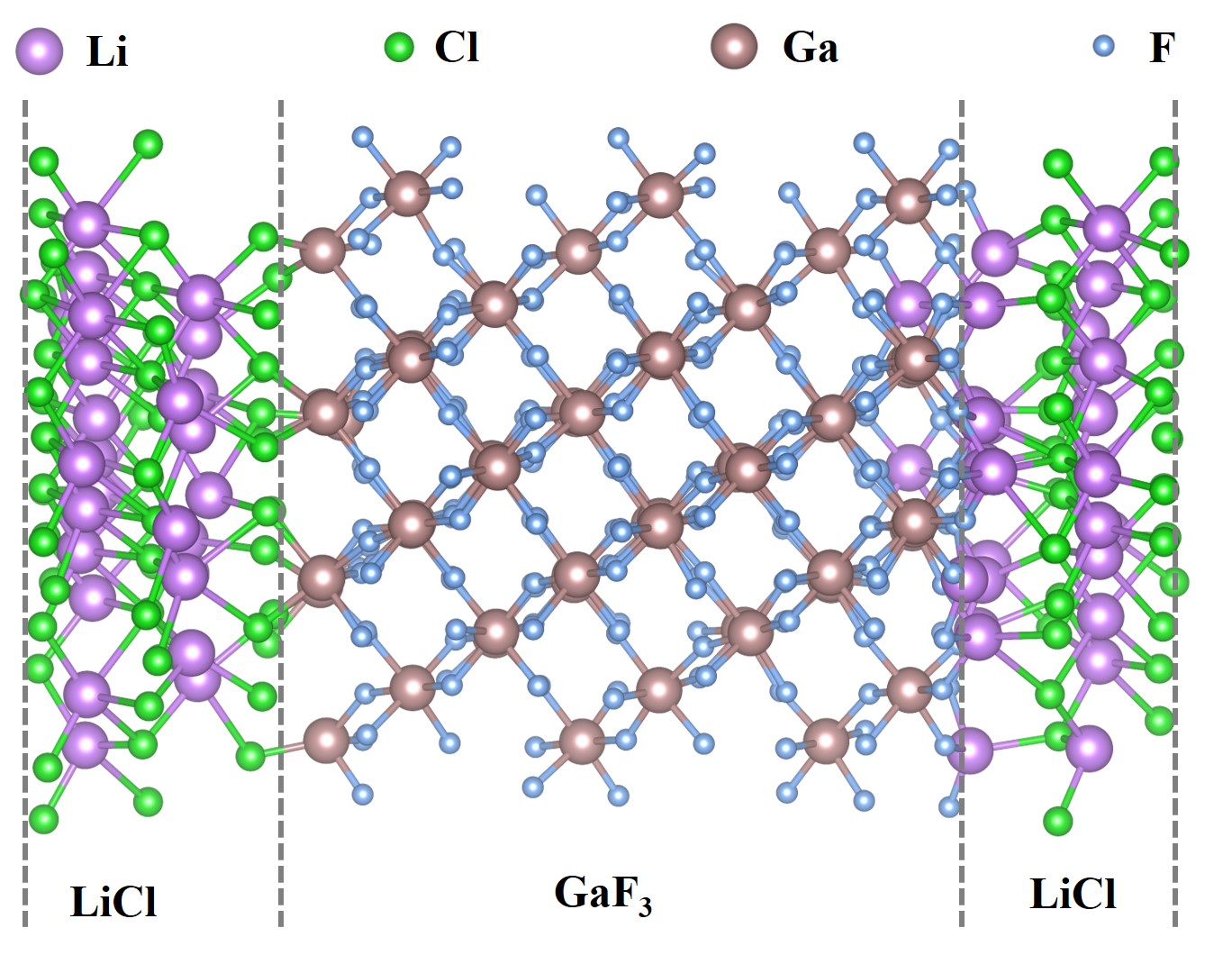}
\caption{Atomic structure of the LiCl(001)/GaF$_3$(001) interface used in this study.}
\label{fig:LiCl-GaF3_RMSE}
\end{figure}

Atomistic modeling of solid–solid interfaces is essential for understanding the synthesizability and stability of materials~\cite{xiao2020understanding}. Such interfaces are notoriously challenging to model because their heterogeneous nature often requires long-period supercells and an accurate treatment of long-range interactions, including charge transfer and polarization, which are typically beyond the scope of conventional MLIPs. To assess whether fine-tuned foundation models can capture these effects, we consider a representative heterointerface from~\cite{kim2024learning}, namely LiCl(001)/GaF$_3$(001). The atomic structure is showin in Figure~\ref{fig:LiCl-GaF3_RMSE}.

Table~\ref{tab:LiClGaF} reports the test RMSEs for energies and forces before and after fine-tuning, together with reference results from~\cite{kim2024learning}. The CACE-LR method, which employs a latent Ewald summation to explicitly describe long-range interactions, already achieves impressive accuracy. Remarkably, however, the fine-tuned foundation model (MACE-FT) performs even better, despite lacking any explicit long-range correction. We attribute this to the fact that large pre-trained models may implicitly encode aspects of long-range physics during pre-training, which fine-tuning then leverages and refines for the target interface. This observation highlights an intriguing direction: foundation models, when fine-tuned, might generalize long-range effects in ways not captured by traditional MLIPs. We plan to further investigate this phenomenon in future work.

\begin{table}[h]
\centering
\caption{\label{tab:LiClGaF} 
Test RMSEs for energies (E, meV/atom) and forces (F, meV/\AA) of the LiCl(001)/GaF$_3$(001) interface. 
Results for CACE-LR are taken from~\cite{kim2024learning}; “--” indicates values not reported. 
The fine-tuned MACE-MPA-0 (MACE-FT) achieves the best accuracy across both metrics.}
{\renewcommand{\arraystretch}{1.3}
\begin{ruledtabular}
\begin{tabular}{lccc}
 & MACE-MPA-0 & CACE-LR & MACE-FT \\
\hline
E (meV/atom) & 97.04 & -- & \textbf{0.09} \\
F (meV/\AA)  & 179.2 & 67.8 & \textbf{18.23} \\
\end{tabular}
\end{ruledtabular}
}
\end{table}

\section{SUMMARY AND PERSPECTIVE}
\label{sec:summary}

This tutorial provides a practical guide to fine-tuning universal machine-learned interatomic potentials (U-MLIPs), using the MACE-MP-0 model as a representative example. We cover the end-to-end process—from dataset preparation and hyperparameter selection to model training and application—with the goal of helping researchers efficiently adapt pre-trained models to their systems of interest. The effectiveness of fine-tuning is demonstrated through representative examples, including solid-state electrolytes, stacking fault defects in metals, surface interactions in low-dimensional materials, and more complicated heterointerfaces. To support reproducibility, all code and data are made publicly available, along with scripts for running molecular dynamics simulations using the RBMD platform.

While this tutorial focuses on MACE-MP-0, the strategies discussed are readily transferable to other U-MLIP frameworks. We hope that this work will help lower the barrier to entry for applying fine-tuned MLIPs in atomistic simulations, and encourage their broader use in both academic and industrial settings.

That said, caution is required when fine-tuning or applying MLIPs. These models approximate the potential energy surface purely from data and do not inherently incorporate physical laws or constraints. Their accuracy is ultimately determined by the coverage and quality of the training dataset. We point out several important considerations and open directions for future development of MLIPs.

\paragraph{Uncertainty Quantification and Active Fine-Tuning.}
Uncertainty quantification (UQ) is essential for evaluating model reliability and guiding data selection in active learning and fine-tuning. Techniques such as ensemble variance, dropout inference, and Bayesian approximations help identify regions of high predictive uncertainty, enabling targeted sampling and noise reduction. In fine-tuning, UQ reduces data redundancy and ensures that added configurations meaningfully enhance model performance. A central challenge is the automated construction of fine-tuning datasets informed by uncertainty, task relevance, or physical constraints. Once such datasets can be systematically generated, selecting or adapting models becomes significantly more efficient. Developing robust pipelines for uncertainty-aware data generation is thus key to scalable, task-adaptive fine-tuning~\cite{Huang2023UncertaintyDistillation}.

\paragraph{Building Intermediate-Sized MLIPs for Specific Tasks.}
Although U-MLIPs demonstrate excellent generalization across a wide range of materials systems, they may not be optimal for small- to medium-sized systems or domain-specific tasks. In such cases, it may be more effective to either fine-tune a compact pre-trained model or train a lightweight architecture from scratch on a focused dataset~\cite{Song2024GeneralMLIP, siddiqui2024machine}. This approach offers a favorable trade-off between interpretability, computational efficiency, and predictive accuracy—particularly when computational resources are limited or when real-time inference is required for large-scale simulations. Lightweight variants of ACE and other descriptor-based models remain valuable tools for such targeted applications.

\paragraph{The Need for Comprehensive Benchmarking Platforms for MLIPs.}
As MLIPs continue to evolve into foundation models with broad applicability, there is an increasing need for benchmarking platforms that go beyond traditional accuracy metrics such as RMSE. Evaluating an MLIP's true usefulness requires not only numerical accuracy, but also physical fidelity, generalization ability, and reliability in practical simulations. Conventional benchmarks often fail to capture these dimensions, making it difficult to assess whether a model is suitable for deployment in real-world applications. To address this gap, recent efforts have begun to develop more comprehensive evaluation frameworks that emphasize model robustness, transferability, and physics-informed behavior. One notable example is \textit{MLIP Arena}~\cite{Chiang2025MLIPArena}, a newly proposed platform that offers architecture-agnostic, task-oriented benchmarking of open-source, open-weight MLIPs. It represents a promising step toward standardized, holistic assessment of model performance in complex materials modeling tasks.

\paragraph{Post-Training Strategies.} After pre-training a foundation model, post-training serves as a crucial step to further adapt and optimize the model for specific downstream applications. One effective post-training technique is \emph{model distillation}, in which a simpler model (the ``student") is trained to replicate the behavior and predictive performance of a larger, more complex model (the ``teacher"—in this case, U-MLIPs). This is typically achieved by using the teacher’s outputs or internal representations as supervisory signals for training the student~\cite{Ekstrom2023Distillation}. Recent work~\cite{Amin2025Distilling} demonstrates that matching the Hessians of energy predictions between the teacher and student enables effective distillation. In particular, large-scale foundation models can be distilled into compact, task-specific MLIPs tailored to a particular chemical subdomain, achieving up to 50$\times$ speedup in inference time compared to the original foundation model. Future investigations on this topic will be conducted using the RBMD platform.

\section{Acknowledgements}
\label{sec:ac}

We gratefully acknowledge the original developers of MACE, particularly Prof. Gábor Csányi's group, for their insightful discussions and valuable contributions. We also appreciate the helpful suggestions and exchanges on the MACE GitHub repository (\url{https://github.com/ACEsuit/mace}) that have informed and improved this tutorial on fine-tuning MACE foundation models. Readers are encouraged to consult the official MACE documentation (\url{https://mace-docs.readthedocs.io/en/latest/}), upon which parts of this tutorial are based. Finally, we gratefully acknowledge the author of the GitHub repository (\url{https://github.com/BingqingCheng/cace-lr-fit}) for generously sharing the dataset that enabled our last two numerical examples.

\section{Codes and Data Availability}
\label{sec:apd:codes}

The datasets and source code supporting all numerical results presented in this work are publicly available in the tutorial repository:
\url{https://github.com/John2021-hub/mace-ft-tutorial.git}.

\section{MACE Architecture}
\label{sec:apd:mace}

In this subsection, we introduce the main idea behind the construction of the MACE architecture. We provide only a brief overview here and refer interested readers to \cite{batatia2022mace} for more detailed constructional insights.

First, a graph is defined by connecting two nodes (atoms) with an edge if they are within each other’s local environment. The local environment, $\mathcal{N}(i)$, consists of all atoms $j$ surrounding a central atom $i$ such that $\|\mathbf{r}_{ij}\| \leq r_{\text{cut}}$, where $\mathbf{r}_{ij}$ is the vector from atom $i$ to atom $j$, and $r_{\text{cut}}$ is a cutoff hyperparameter. The feature vector of node $i$ is denoted by $\mathbf{h}_i^{(t)}$, expressed in the spherical harmonic basis, with indices $l$ and $m$. The superscript $t$ represents the iteration step (analogous to the ``layers” in graph neural networks’ message-passing). These node features, $\mathbf{h}_i^{(t)}$, reflect the chemical environment of the atoms.

It is worth noting that all MACE models consist of only two layers. Due to the specific construction, MACE can be regarded as a higher-order message-passing neural network. This enables it to achieve comparable expressiveness with just two layers, which is a key strength of the MACE model.

The node features on 0-th layer denoted as $\mathbf{h}_i^{(0)}$ are initialized as a (learnable) embedding of the chemical elements with atomic numbers $z_i$ into $k$ channels:
\begin{equation}
h_{i,k00}^{(0)} = \sum_{z} W_{kz} \delta_{z z_i}.
\end{equation}
This type of mapping has been widely applied to graph neural networks~\cite{schutt2017schnet, schutt2021equivariant} and other models~\cite{willatt2018feature, darby2023tensor}.

Next, for each atom, the features of its neighbors are combined with the interatomic displacement vectors, $\mathbf{r}_{ij}$, to form the one-particle basis $\phi_{ij,knl_1l_2m_2}^{(t)}$. The construction of MACE employs the idea from the ACE architecture, which is a natural many-body representation for capturing the symmetry of many-body particle systems. The radial distance $r_{ij}$ is used as an input into a learnable radial function $R(r_{ij})$, with several outputs that correspond to different ways in which the displacement vector and node features can be combined while maintaining equivariance~\cite{wigner2012group}:
\begin{align}
\phi_{ij,knl_1l_2m_3}^{(t)} = \sum_{l_1m_1l_2m_2} & C_{\eta_1,l_1m_1,l_2m_2}^{l_3m_3} R_{knl_1l_2l_3}^{(t)}(r_{ij}) \nonumber \\
&Y_{l_1m_1}^{m_1} (\hat{r}_{ij}) h_{j,kl_2m_2}^{(t)},
\end{align}
where $Y_{l}^{m}$ are the spherical harmonics, and $C_{\eta_1,l_1m_1,l_2m_2}^{l_3m_3}$ denotes the Clebsch-Gordan coefficients. There are multiple ways of constructing an equivariant combination with a given symmetry $(l_3, m_3)$, and these multiplicities are enumerated by the path index $\eta_1$~\cite{ACECompleteness}.

The one-particle basis $\phi$ is summed over the neighborhood and expanded as a linear combination of $k$ channels with learnable weights to form the permutation-invariant atomic basis $A_i$:
\begin{equation}
    A_{i,kl3m3}^{(t)} = \sum_{\tilde{k},\eta_1} W_{\tilde{k}\tilde{k}k\eta_1 l_3} \sum_{j \in \mathcal{N}(i)} \phi_{ij,\tilde{k}\eta_1 l_3m_3}^{(t)}.
\end{equation}

Higher-order (many-body) symmetric features are created on each atom by taking products of the atomic basis, $A$, with itself $\nu$ times, resulting in the ``product basis". The product basis is then contracted with the generalised Clebsch-Gordan coefficients $C_{\eta_\nu lm}^{LM}$ to obtain the equivariant higher-order basis, $B_i$ \cite{ACECompleteness}:
\begin{equation}
    B_{i,\eta_\nu kLM}^{(t),\nu} = \sum_{lm} C_{\eta_\nu lm}^{LM} \prod_{\xi=1}^{\nu} A_{i,kl_\xi m_\xi}^{(t)},
\end{equation}
where bold $lm$ denotes the $\nu$-tuple of $l$ and $m$ values and similarly to Equation (2), $\eta_\nu$ enumerates the number of possible couplings to create the features with equivariance $LM$.

Finally, a ``message" $m_i$ is created on each atom as a learnable linear combination of the equivariant many-body features:
\begin{equation}
    m_{i,kLM}^{(t)} = \sum_{\nu} \sum_{\eta_\nu} W_{zi\eta_\nu kLM}^{(t),\nu} B_{i,\eta_\nu kLM}^{(t),\nu}.
\end{equation}

The recursive update of the node features $(t : 0 \rightarrow 2)$ is obtained by adding the message to the atoms' features from the previous iteration, with weights that depend on the chemical element ($z_i$) that are also responsible for the mixing of the chemical embedding $k$ channels:
\begin{equation}
    h_{i,kLM}^{(t+1)} = \sum_{\tilde{k}} W_{kL,\tilde{k}}^{(t)} m_{i,\tilde{k}LM}^{(t)} + \sum_{\tilde{k}} W_{kziL,\tilde{k}}^{(t)} h_{i,\tilde{k}LM}^{(t)}.
\end{equation}

The site energy is a sum of read-out functions applied to node features from the first and second layers. The read-out function is defined as a linear combination of rotationally invariant node features for the first layer, and as a multi-layer perceptron (MLP) for the second layer.
\begin{equation}
    E_i = \sum_{t=1}^{2} E_i^{(t)} = \sum_{t=1}^{2} \mathcal{R}^{(t)}\left( h_i^{(t)} \right),
\end{equation}
where
\begin{equation}
    \mathcal{R}^{(t)}\left( h_i^{(t)} \right) =
    \begin{cases} 
        \sum_k W_k^{(t)} h_{i,k00}^{(t)} & \text{for } t = 1 \\ 
        \text{MLP} \left( \left\{ h_{i,k00}^{(t)} \right\}_k \right) & \text{for } t = 2 
    \end{cases}.
\end{equation}

The forces and stresses on the atoms are calculated by taking analytical derivatives of the total potential energy with respect to the positions of the atoms.

\bibliography{apssamp}

\end{document}